\documentstyle[pra,aps,epsfig,multicol]{revtex}

\begin{document}
\title{\Large\bf Raman scattering in ${\rm C}_{60}$ and 
       ${\rm C}_{48}{\rm N}_{12}$ aza-fullerene:  
       First-principles study}
\author{ Rui-Hua Xie and Garnett W. Bryant} 
\address{National Institute of 
Standards and Technology, Gaithersburg, MD 20899-8423, USA}
\author{Vedene H. Smith, Jr.}
\address{Department of Chemistry, Queen's University, Kingston,
ON K7L 3N6, Canada}
\date{\today}
\maketitle

\begin{abstract}
 
We carry out large scale {\sl ab initio} calculations of Raman  scattering
activities and Raman-active frequencies (RAFs) in 
${\rm C}_{48}{\rm N}_{12}$ aza-fullerene. The results are compared 
with those of ${\rm C}_{60}$. Twenty-nine  non-degenerate polarized and 29
doubly-degenerate unpolarized RAFs are predicted for
${\rm C}_{48}{\rm N}_{12}$. The RAF of the strongest Raman signal in
the low- and high-frequency regions  and the lowest
and highest RAFs for ${\rm C}_{48}{\rm N}_{12}$ are
almost the same as those of ${\rm C}_{60}$. The study of ${\rm C}_{60}$
reveals the importance of electron correlations and the choice of
basis sets in the {\sl ab initio} calculations. Our best calculated results for
${\rm C}_{60}$ with the B3LYP hybrid density functional theory  are
in excellent agreement with  experiment and demonstrate the desirable
efficiency and accuracy of this theory for obtaining
quantitative information on the vibrational properties of these
molecules.
 
{\bf PACS (numbers):} 36.20.Ng, 31.15.Ar, 78.30.-j, 78.30.Na, 82.80.Gk
 
\end{abstract}

\begin{multicols}{2}

\section{Introduction}

In 1985, ${\rm C}_{60}$, a fascinating molecule formed as a 
truncated icosahedron with 20 hexagonal and 12 pentagonal faces, and 60 vertices, 
 was discovered by Kroto {\sl et al.}\cite{kroto}.  Since then,  a 
third form of pure carbon, called {\sl fullerenes}\cite{curl91},  
has been extensively studied. This kind of molecule can crystallize in a variety of three-dimensional 
structures\cite{wk90a,drh91}, with  an even number of 
three-coordinated ${\rm sp}^{2}$ carbon atoms that arrange themselves 
into 12 pentagonal faces and any number ($> 1$) of hexagonal 
faces\cite{curl91}. The macroscopic synthesis of  soot\cite{wk90a}, 
which contains ${\rm C}_{60}$ and other fullerenes in large compounds, 
plus the straightforward purification techniques, which make pure fullerene 
materials available,  opened new research opportunities in science, 
engineering and technology\cite{book1,book3,book4,xie1,book5}.  
In the meantime, doped fullerenes have also  attracted a great deal of 
researchers' interest due to their remarkable structural, electronic, 
optical and magnetic properties\cite{book1,book3,book4,xie1,book5}. 
For example,  the doped tubular fullerenes can 
exhibit large third-order optical nonlinearities\cite{xie1} and be ideal candidates 
as photonic devices\cite{book4} for  all-optical switching, data processing, 
eye and sensor protection (optical limiter). Another example is 
alkali metal-doped $C_{60}$ crystals, which can be superconducting
\cite{book1}. Because of their unique structure and electronic properties\cite{book1}, 
fullerenes can be doped in several different 
ways including endohedral doping\cite{book1,jacs02}, substitutional doping 
\cite{guo91,hummelen95,andreoni96a,hultman01,stafstrom,xiecpl02}, and exohedral doping 
\cite{book1,book3,book4,book5}. 

Since the average carbon-carbon bond length in 
${\rm C}_{60}$ is slightly larger than that in graphite, which  can 
only be doped by boron, and the force constants \cite{book1} 
are somewhat weakened by the curvature of the ${\rm C}_{60}$ surface, 
both boron and nitrogen can substitute for one or more carbon atoms in 
${\rm C}_{60}$\cite{guo91,hummelen95,andreoni96a,hultman01,stafstrom,xiecpl02}.  
In 1991, Smalley and coworkers\cite{guo91} successfully synthesized boron-substituted 
fullerenes ${\rm C}_{60-n}{\rm B}_{n}$ with $n$ between 1 and at least 6. 
In 1995, Hummelen {\sl et al.}\cite{hummelen95} reported a very efficient method 
of synthesizing ${\rm C}_{59}{\rm N}$, which has led to a number of detailed studies of the 
physical and chemical properties of ${\rm C}_{59}{\rm N}$
\cite{book3,book4,book5,andreoni96a}. Recently, Hultman {\sl et al.}\cite{hultman01} 
have synthesized nitrogen-substituted derivatives of ${\rm C}_{60}$ with more than one 
nitrogen atom and  reported the existence of a novel ${\rm C}_{48}{\rm N}_{12}$ aza-fullerene
\cite{hultman01,stafstrom}. Very recently, we have studied the bonding, electronic structure, 
Mulliken charge, infrared (IR) spectrum, and NMR  of ${\rm C}_{48}{\rm N}_{12}$ by using 
density functional theory (DFT) and the 6-31G basis set\cite{xiecpl02}. We characterized 
58 IR spectral lines, eight  $^{13}{\rm C}$ and two $^{15}{\rm N}$ NMR spectral signals
 of ${\rm C}_{48}{\rm N}_{12}$, and  demonstrated that this aza-fullerene has potential 
applications as semiconductor components for nanometer electronics because of its small 
energy gap, as a promising electron donor for molecular electronics, and as a good diamagnetic 
material because of the enhancement of diamagnetic factors in the carbon atoms\cite{xiecpl02}.  
The characterization of ${\rm C}_{48}{\rm N}_{12}$ is a timely problem both from the viewpoint 
of its practical applications and to understand  doping in ${\rm C}_{60}$ derivatives. In the 
present paper, we characterize the Raman spectrum of ${\rm C}_{48}{\rm N}_{12}$. 

As a  material is doped with foreign atoms, its mechanical, 
electronic, magnetic and optical properties  change 
\cite{book1,book4}.  The ability to control such induced changes 
is vital to progress in material science.  Raman  and IR 
spectroscopic techniques\cite{book8a,book8b} are basic, useful experimental 
tools  to investigate how  doping modifies the structural and dynamical 
properties  of the pristine material and to understand the physical 
origin of such induced changes. Over the past 10 years, both techniques 
have been widely used to study the vibrational properties of ${\rm C}_{60}$ 
\cite{ir1,ir2,ir3,ir4,rm1,rm2}, its derivative compounds
\cite{d1,d2}, and (doped) carbon nanotubes\cite{rao97a}. 
It has been  shown that ${\rm C}_{60}$ has  in total 
46 vibrational modes including 4 IR-active \cite{ir1,ir2,ir3,ir4} 
and 10 Raman-active \cite{rm1,rm2} vibrational modes. Well-resolved 
Raman spectra\cite{d1} are also available for ${\rm C}_{60}$ and a 
number of its derivative compounds. 

In this paper, we perform first-principles calculations of Raman scattering activities 
(RSAs) and Raman-active frequencies (RAFs)  in both ${\rm C}_{48}{\rm N}_{12}$  and ${\rm C}_{60}$ 
using density functional theory (DFT) and the restricted Hartree-Fock (RHF) method.
Very recently, Choi {\sl et al.}\cite{choi00a} have theoretically assigned 
all 46 vibrational modes of ${\rm C}_{60}$, including a scaling of the force 
field by using  Pulay's method. In this paper,  however, 
we  carry out {\sl ab initio} a series of calculations for ${\rm C}_{60}$. We 
want to test the efficiency and accuracy of such 
first-principles calculations and study, in detail, 
basis set effects on RSAs and RAFs by comparing our theoretical results 
with available experiments\cite{book1,rm1,rm2,d1,d2}. To the best of our knowledge, 
such basis set effects on the RSAs and RAFs of ${\rm C}_{60}$ have not been considered 
before. These calculations for ${\rm C}_{60}$ give us a benchmark for assigning the Raman-active vibrational 
modes of ${\rm C}_{48}{\rm N}_{12}$  and provide us with constructive insight into the microscopic mechanisms 
responsible for the difference between ${\rm C}_{48}{\rm N}_{12}$ and ${\rm C}_{60}$. 
 We find that the 10 RAFs for ${\rm C}_{60}$, obtained  
by using a hybrid DFT method and large basis sets, are in excellent agreement 
with Raman experiments. We predict that ${\rm C}_{48}{\rm N}_{12}$ aza-fullerene has 
58 RAFs including 29 non-degenerate polarized modes and 29 doubly-degenerate unpolarized 
modes. 

This paper is organized as follows. Section II briefly reviews the 
{\sl ab initio} methods, basis sets  and the theory of calculating Raman scattering 
activities. Section III presents our Raman results for both 
${\rm C}_{60}$ and ${\rm C}_{48}{\rm N}_{12}$ obtained by using RHF and DFT 
methods, and  the results are compared to results 
obtained by other theoretical methods. Our conclusions are given in  section IV. 

\section{Theory}

\subsection{{\sl Ab Initio} Methods and Basis Sets} 

{\sl Ab initio} methods obtain information by solving variationally the
Schr\"{o}dinger equation without fitting parameters to experimental
data. Instead,  experimental data guides the selection of the {\sl ab initio}
methods rather than directly entering the computational procedures.
Fullerenes have been challenging molecules for {\sl ab initio} calculations  because of
their size\cite{jc95}. Recent advances in {\sl ab initio} electronic
structure methods and parallel computing have brought a substantial
improvement in the capabilities  to predict and study 
the properties of large molecules. The coupled cluster method\cite{cook98}
 has been used to predict phenomena in $C_{20}$\cite{taylor95}. Other
{\sl ab initio} methods, which are less
demanding in terms of computation cost than the coupled cluster
method, have been used for much larger fullerenes and carbon
nanotubes, for example, ${\rm C}_{60}$
\cite{disch86,haser91}  with self-consistent field  and 
Moller-Plesset second-order (MP2) theory,
${\rm C}_{240}$ \cite{bakowies95} and carbon nanotubes\cite{jc02} with
density functional theory (DFT), and ${\rm C}_{540}$ \cite{scuseria95} 
with the Hartree-Fock (HF) method. The major expense in HF and DFT 
calculations arises from solving the electronic quantum Coulomb problem. 
The effective Hamiltonian diagonalization (a procedure that scales as 
$N_{b}^{3}$, $N_{b}$ being the number of basis functions) represents 
only a minor portion of the computational time in calculations of 
molecular clusters containing up to several hundred atoms \cite{strout95}. 
In addition, DFT \cite{kohn64}  requires an additional three-dimensional numerical 
quadrature to obtain the exchange and correlation energies\cite{cook98}.
Both HF and DFT methods have been implemented into the Gaussian 98
program\cite{gaussian}. In this paper, all calculations are performed by 
using the Gaussian 98 program\cite{gaussian,nist}. 
 
 One of the approaches  inherent in all {\sl ab initio} methods is the introduction of a basis set\cite{cook98}.
If the basis set is complete, exact expansions of the molecular orbitals can be obtained.
However, a complete basis set  requires  an infinite number of functions, 
which is impossible in actual calculations. Generally, a smaller basis set provides 
a poorer representation. Moreover, only the parts of the molecular orbital 
which correspond to the selected basis can be represented.  Since the computational 
effort of {\sl ab initio} methods scales formally as 
$N_{b}^{4}$\cite{cook98}, it is  important  to make the basis set as small
as possible without compromising the accuracy or missing part of the state space which 
should be represented. Hence,  one purpose of this 
paper  is to determine  the effects of basis sets\cite{gaussian} for the calculations of 
Raman spectra by consdering the Slater-type basis set STO-3G and split valence basis sets 3-21G, 6-31G and  
6-31G(d) (i.e., 6-31G*).

\subsection{Vibrational Analysis and Raman Scattering Activity}

The vibrational analysis of polyatoms described by Wilson {\sl et al.}\cite{ir16} has 
been implemented in the Gaussian 98 program. This analysis 
is valid only  when the first derivatives of the energy with respect to 
the displacement of the atoms are zero (in other words, the geometry used for vibrational 
analysis must be optimized at the same level of theory and with the same basis set 
that the second derivatives are generated with). The force constant matrix ${\bf K}$ is defined as the 
second partial derivatives of the potential $V$ with respect to the displacement 
 of the atoms in cartesian coordinates (for example, $\Delta x_{k}$, $\Delta y_{k}$, $\Delta z_{k}$ 
of the $k$th atom), i.e., a $3n\times 3n$ matrix (n is the number of atoms) whose 
elements are given by 
\begin{equation}
K_{ij} = \left (\frac{\partial^{2} V}{\partial\eta_{i}\partial\eta_{j}}\right)_{0} 
\hspace{1cm} (i,j = 1, 2, ..., 3n)
\end{equation}
where $(\eta_{1}, \eta_{2}, \eta_{3})\equiv (\Delta x_{1}, \Delta y_{1}, \Delta z_{1})$. 
$(...)_{0}$ means that the second partial derivatives are taken at the equilibrium positions 
of the atoms. Usually, the matrix ${\bf K}$ in cartesian coordinates  is 
 converted to a new matrix  $\tilde{{\bf K}}$ in 
mass-weighted cartesian coordinates $q_{i}=m_{i}^{1/2}\eta_{i}$, i.e., 
\begin{equation}
\tilde{K_{ij}}\equiv (m_{i}m_{j})^{-1/2}K_{ij} = 
\left (\frac{\partial^{2} V}{\partial q_{i} \partial q_{j}}\right)_{0},
\end{equation}
where $m_{i}$ is the mass of the atom. Then, the eigenvalues $\lambda_{k}$ of  $\tilde{{\bf K}}$ 
give the fundamental frequencies $\tilde{\nu_{k}}$ (in the unit of ${\rm cm}^{-1}$), i.e., 
 $\tilde{\nu_{k}}=\lambda_{k}^{1/2}/(2\pi c)$, where $c$ is the velocity of light in vacuum. 
The eigenvectors give the normal modes.  

To lowest order,  Raman intensities are proportional to the derivatives of the dipole 
polarizability with respect to the vibrational normal modes of the material, evaluated at 
the equilibrium geometry. For example,  most Raman scattering experiments use  a plane-polarized 
incident laser beam. The direction of the incident beam, the polarization
direction of this beam and the direction of observation are chosen to be 
perpendicular to each other. Under these circumstances, the first-order differential cross
section for Raman scattering in the $q$th vibrational mode is written as\cite{ir17}
\begin{equation}
\left(\frac{d\sigma}{d\Omega}\right)_{q}
=\frac{\hbar\omega_{s}^{4}}{90c^{4}\pi\omega_{q}(1-n_{q})}I_{raman}.
\end{equation}
${\rm I}_{\rm raman}$ is the Raman scattering activity, and 
 $n_{q} =\exp{[-\hbar\omega_{q}/(\kappa_{B} T)]}$. 
 $\omega_{s}$ is the frequency of the  scattered radiation, 
$\omega_{q}$ is the frequency of the $q$th vibrational mode, $\kappa_{B}$ is  
Boltzmann's constant, and $T$ is temperature. For the 
special case of $\pi/2$ scattering geometry, ${\rm I}_{\rm raman}$ can be written 
as 
\begin{eqnarray}
I_{raman} &=&5\left(\frac{d\alpha_{xx}}{d\Xi_{q}} +\frac{d\alpha_{yy}}{d\Xi_{q}}+\frac{d\alpha_{zz}}{d\Xi_{q}}\right)^{2}\nonumber\\ 
&+&\frac{7}{4}\left[\left(\frac{d\alpha_{xx}}{d\Xi_{q}}-\frac{d\alpha_{yy}}{d\Xi_{q}}\right)^{2}
+\left(\frac{d\alpha_{xx}}{d\Xi_{q}}-\frac{d\alpha_{zz}}{d\Xi_{q}}\right)^{2}\right.\nonumber\\
&+&\left(\frac{d\alpha_{yy}}{d\Xi_{q}}-\frac{d\alpha_{zz}}{d\Xi_{q}}\right)^{2}
+6\left(\frac{d\alpha_{xy}}{d\Xi_{q}}\right)^{2}\nonumber\\ 
&+&\left. 6\left(\frac{d\alpha_{xz}}{d\Xi_{q}}\right)^{2}+6\left(\frac{d\alpha_{yz}}{d\Xi_{q}}\right)^{2}\right]^{2}, 
\end{eqnarray}
where $\Xi_{q}$ is the normal coordinate corresponding to the $q$th vibrational mode 
and  $\alpha$ is the dipole polarizability tensor. 
 
To obtain the  Raman activities, one must compute the derivatives
of the polarizability with respect to the normal mode
coordinates $\Xi_{q}$. These can be viewed as
directional derivatives in the space of 3N nuclear coordinates and are expressed
in terms of derivatives with respect to atomic coordinates, $R_{k}$. For 
the polarizability component $\alpha_{ij}$ (i, j = x, y, z), we have 
\begin{equation}
\frac{d\alpha_{ij}}{d\Xi_{q}}=\sum_{k=1}^{3N}\frac{\partial\alpha_{ij}}{\partial R_{k}}\xi_{kq}, 
\end{equation}
where $\xi_{kq}=\partial R_{k}/\partial\Xi_{q}$ is the $k$th
atomic displacement of the $q$th normal mode. 
Then, the necessary derivatives can be expressed in terms of the atomic forces
as follows\cite{ir16,ir17}
\begin{eqnarray}
\frac{\partial\alpha_{ij}}{\partial R_{k}}&=&
-\frac{\partial^{3}E}{\partial G_{i}\partial G_{j}\partial R_{k}}=
\frac{\partial^{2}F_{k}}{\partial G_{i}\partial G_{j}},
\end{eqnarray}
where $E$ is the total energy, $G_{i}$ is the $i$th component
of an assumed external electric field ${\bf G}$, and
$F_{k}=\partial E/\partial R_{k}$ is the calculated force 
on the $k$th atomic coordinate.

\section{Results}

\subsection{Geometry Optimizations} 

In our previous work\cite{xiecpl02}, we calculated the bond lengths of 
${\rm C}_{48}{\rm N}_{12}$ by using the B3LYP\cite{becke93} hybrid DFT method 
with the 6-31G basis set. The geometry of ${\rm C}_{48}{\rm N}_{12}$ is shown in Fig.1, where the 
10 unique sites (vertices 1 to 10) can be identified from nuclear magnetic resonance\cite{xiecpl02}. 
It was found that there are 15 unique types of bonds in ${\rm C}_{48}{\rm N}_{12}$\cite{xiecpl02}: 6 
nitrogen-carbon bonds and 9 carbon-carbon bonds. In contrast, ${\rm C}_{60}$ has 
one kind of single (C-C) bond and one kind of double bond (C=C).

\begin{center}
\epsfig{file=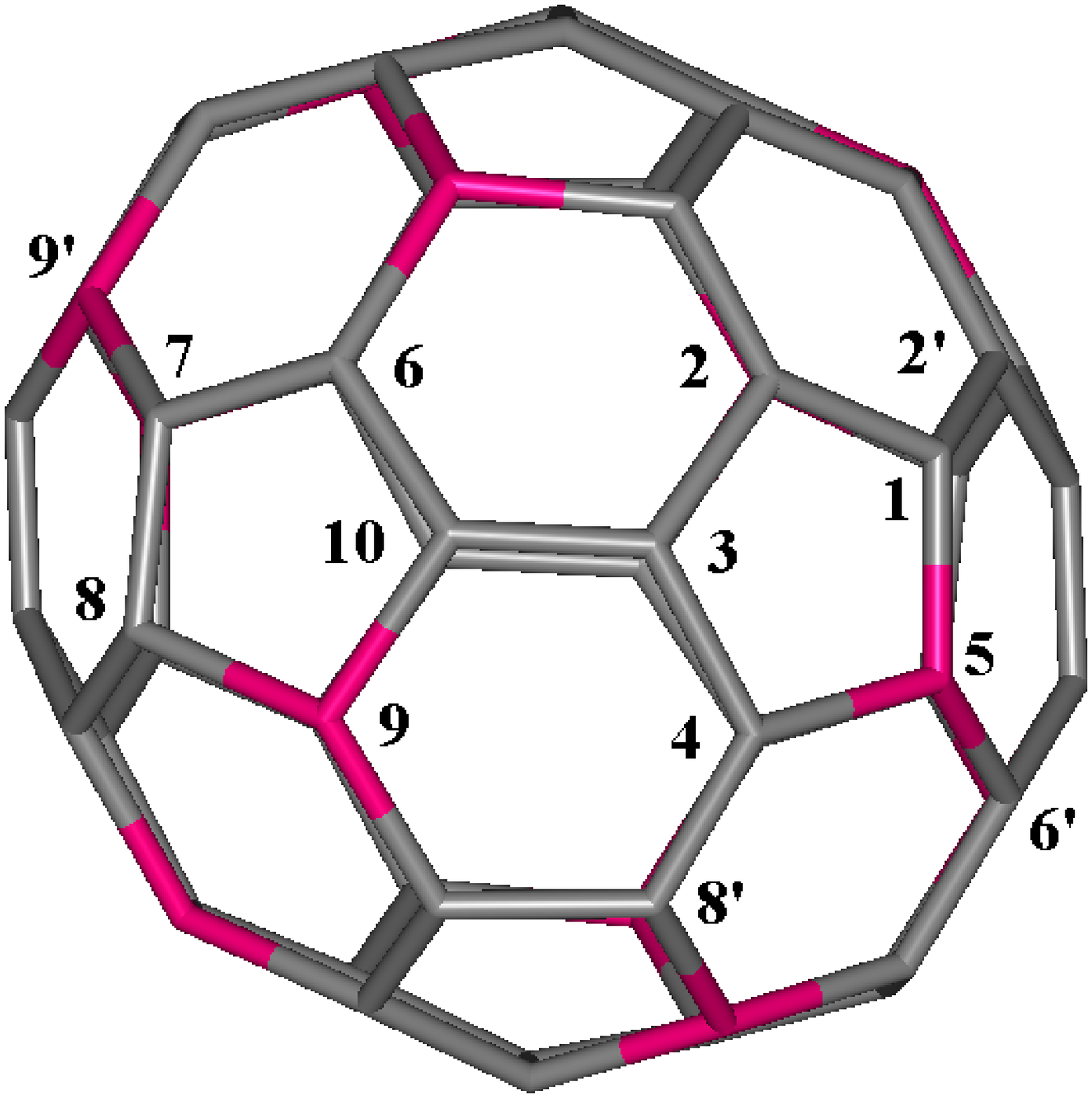,width=6cm,height=6cm}
\end{center}

\begin{quote}
{\bf  FIG. 1.}:  {\small ${\rm C}_{48}{\rm N}_{12}$ structure. Red  is for nitrogen  sites and grey
for carbon sites. There are 10 unique vertices (1 to 10) as labeled. The 15 bonds between
labeled vertices are all unique.}
\end{quote}

In this paper, we discuss the effect of basis sets using different {\sl ab initio} methods. 
Before calculating the RSAs and RAFs of both ${\rm C}_{60}$ and ${\rm C}_{48}{\rm N}_{12}$, 
we  first optimize their geometries  by using  RHF and B3LYP hybrid DFT methods 
with a variety of basis sets including STO-3G, 3-21G, 6-31G and 6-31G*. The  B3LYP DFT 
method includes a mixture of Hartree-Fock (exact) exchange, Slater local exchange\cite{slater51}, 
Becke 88 non-local exchange\cite{becke88}, the VWN III local exchange-correlation functional
\cite{vosko80} and the LYP correlation functional\cite{lyp88}. The optimized bond lengths 
for  ${\rm C}_{48}{\rm N}_{12}$ and ${\rm C}_{60}$ are listed in Table I.  
Compared to ${\rm C}_{60}$, in ${\rm C}_{48}{\rm N}_{12}$ the lengths of bonds 
that were ${\rm C}_{60}$ double bonds (for example, C1-C$2^{'}$, C3-C10) increase, 
while the lengths of bonds that were  single bonds in ${\rm C}_{60}$ 
(for example, C3-C4, C6-C10) decrease,  yielding two bonds with a small difference in length. 
According to our B3LYP/6-31G* 
calculation, the bond length difference is reduced from 0.059 \AA \ in 
${\rm C}_{60}$ to 0.027-0.041 \AA \  in ${\rm C}_{48}{\rm N}_{12}$. This 
can be explained by the fact that the 12 extra electrons in ${\rm C}_{48}{\rm 
N}_{12}$ fill $t_{1u}$ and $t_{1g}$ orbitals which have some antibonding 
character and consequently increase the length of C=C double bonds. 
Compared with B3LYP calculations  and experiments 
\cite{c60exp} for ${\rm C}_{60}$, RHF generally underestimates the lengths 
of C=C and C-C bonds by about 0.5\%. Judging from the success of the B3LYP geometries for ${\rm C}_{60}$,  
we conclude that the role of electron correlation is important in the 
accurate description of localization/delocalization of $\pi$ electrons 
in ${\rm C}_{60}$. Such bond equalization has also been observed in the DFT 
calculations for ${\rm K}_{3}{\rm C}_{60}$ by Bohnen {\sl et al.} 
\cite{bohnen95} and for ${\rm C}_{60}^{-6}$ by Choi {\sl et al.} 
\cite{choi00a}. To prevent the collapse of the valence functions into 
the inner shell and ensure an adequate description of bonding interactions which 
involve overlap of valence funcitons, the 6-31G* basis set for 
carbon and nitrogen provides a better description of the inner-shell region as well as 
the valence region than STO-3G, 3-21G and 6-31G basis sets\cite{gaussian,basis00}.  Overall, the bond length 
${\rm L}$ in ${\rm C}_{60}$ or ${\rm C}_{48}{\rm N}_{12}$, as listed in Table I, follows the order: 
${\rm L}_{6-31G*} < {\rm L}_{3-21G} < {\rm L}_{6-31G}< {\rm L}_{STO-3G}$, 
where  ${\rm L}_{6-31G*}$ for ${\rm C}_{60}$ is in good agreement with experiment. 
The ordering of bond lengths suggest that increasing the flexibility of the 
basis from 3-21 to 6-31G favors expansion of the ${\rm C}_{60}$, while the extra polarization functions 
in 6-31G* are needed to recontract ${\rm C}_{60}$. As a result, the 3-21G basis set 
fortuitously gives bond lengths closer to experiment than does the bigger 6-31G basis 
set. 

\end{multicols}

\noindent
{\bf Table I:}   Bond lengths (${\rm L}$, in ${\rm\AA}$ )
in ${\rm C}_{48}{\rm N}_{12}$ and ${\rm C}_{60}$ calculated by using B3LYP  hybrid DFT  and RHF methods
with STO-3G, 3-21G, 6-31G and 6-31G* basis sets. $i$ and $j$ in
C$i$ and N$j$ are the site numbers labeled in Fig.1. The symbols ``C-C"
and ``C=C" in parenthesis denote the original single and double bonds
in ${\rm C}_{60}$. The averaged bond length ${\rm L}_{ave} = ({\rm L}_{C=C}+2{\rm L}_{C-C})/3$.
\begin{center}
\begin{tabular}{lcccccccccc}\hline\hline
 &      &\multicolumn{2}{c}{\underline{STO-3G}}& \multicolumn{2}{c}{\underline{\ \ 3-21G\ \ \ }}
 &\multicolumn{2}{c}{\underline{\ \ \ 6-31G\ \ \ }} & \multicolumn{2}{c}{\underline{\ \ \ 6-31G*\ \ }} &\underline{\ \ \ EXP.\ \ \  } \\
{\small Fullerene}& {\small Bond} &   ${\small {\rm L}_{dft}}$  &  ${\small {\rm L}_{rhf}}$ & ${\small
{\rm L}_{dft}}$  & ${\small {\rm L}_{rhf}}$ &  ${\small {\rm L}_{dft}}$  & ${\small {\rm L}_{rhf}}$
& ${\small  {\rm L}_{dft}}$  & ${\small {\rm L}_{rhf}}$ & {\small Ref.\cite{c60exp} } \\ \hline
${\rm C}_{48}{\rm N}_{12}$  & C1-C2$^{'}$ (C=C) &1.437   &1.415 &1.417 &1.408  & 1.422 &1.413 &1.416  & 1.402 & \\
& C1-C2  (C-C) &1.428  &1.391  &1.410 &1.386  & 1.413 &1.388  &1.406   & 1.384 &  \\
& C1-N5 (C-C) &1.474  &1.460 &1.432  &1.421 & 1.432 &1.416 &1.430  &1.427 & \\
& C3-C2 (C-C) &1.470 &1.466  &1.452 &1.452 & 1.449 &1.448&1.446 &1.452 &  \\
& C3-C4 (C-C)  &1.409  &1.359  &1.395 &1.362  & 1.397 &1.365  &1.390  & 1.359 &  \\
& C3-C10  (C=C) &1.456  &1.457  &1.432 &1.435 & 1.434(6)  &1.437 &1.431(4)   &1.436(8)&  \\
& C4-C8$^{'}$ (C=C) &1.457 &1.459  &1.433 &1.437 & 1.435(4) & 1.439 &1.431(3) & 1.437(2) & \\
& C4-N5 (C-C) &1.471  &1.455  &1.431  &1.413  & 1.429 &1.411  &1.422  &1.405&  \\
& N5-C6$^{'}$ (C=C) &1.475  &1.464  &1.430  &1.429 & 1.432  &1.428  &1.429  &1.431 &  \\
& C6-C10 (C-C)&1.412   &1.364  &1.396 &1.366 &1.400 &1.371  &1.394   &1.363   & \\
& C6-C7 (C-C) &1.440 &1.435  &1.421 &1.418 & 1.422 &1.419  &1.414 &  1.414 & \\
& C7-C8 (C-C)  &1.420  &1.371  &1.402 & 1.372 & 1.407  &1.376 &1.402   &1.370 &  \\
& C7-N9$^{'}$ (C=C) &1.459 &1.442 &1.416  &1.405 & 1.419 &1.407 &1.410 & 1.396 &  \\
& C8-N9 (C-C) &1.451 &1.431  &1.418 &1.408  & 1.415 &1.401 &1.408 &1.406 &  \\
& N9-C10 (C-C) &1.461  &1.436  &1.427 & 1.404  & 1.423 &1.399 &1.413  &1.392 &  \\
${\rm C}_{60}$ & C=C (C1-C$2^{'}$) &1.413  &1.376  &1.390 &1.367 & 1.398  &1.375 &1.395    &1.373 &1.391  \\
 & C-C (C1-C2, C1-C5) &1.477  &1.463 &1.460 &1.453 & 1.459  &1.452 &1.454    &1.449 & 1.455 \\
 & ${\rm L}_{ave}$ & 1.456 & 1.434 & 1.437 & 1.424 & 1.439 & 1.426 & 1.434 & 1.424 & 1.434 \\ \hline
\end{tabular}
\end{center}

\begin{multicols}{2}
According to our DFT or RHF calculations, the diameter of ${\rm C}_{48}{\rm N}_{12}$ 
varies, as compared to that of ${\rm C}_{60}$, from -0.3 \AA \ to +0.2 \AA, which 
has a strong effect on the force constant that determines the frequencies of the 
vibrational modes of molecule. Hence, we expect that the vibrational frequencies of 
${\rm C}_{48}{\rm N}_{12}$ will be strongly influenced by the bond equalization 
effects and the expansion and contraction of the ball. In the following, we 
investigate the vibrational properties of ${\rm C}_{48}{\rm N}_{12}$ 
and ${\rm C}_{60}$ based on the optimized geometries.    
 
\subsection{Raman-active Vibrational Frequencies}
 
Using the Gausian 98 program\cite{gaussian,nist}, we calculated the RAFs of
 both ${\rm C}_{60}$ and ${\rm C}_{48}{\rm N}_{12}$. We compare with 
${\rm C}_{60}$ experiment to determine  the effects of electron correlations 
and basis sets. It should be mentioned that our frequencies have not been
scaled.
 
Table II summarizes the calculated results  for ${\rm C}_{60}$ obtained by
using RHF and B3LYP hybrid DFT methods. As shown by Dresselhaus
{\sl et al.}\cite{dress92}, there are 46 different vibrational modes in the 174
independent normal vibrations of ${\rm C}_{60}$. These modes are 
classified in even and odd parities and in the ten irreducible representatives
of the $I_{h}$ point group\cite{dress92}: the \{$a_{g}$, $a_{u}$\}, \{$t_{1g}$, $t_{1u}$,
$t_{2g}$, $t_{2u}$\}, \{$g_{g}$, $g_{u}$\}  and \{$h_{g}$, $h_{u}$\} modes
are non-, threefold-,  fourfold- and fivefold-degenerate, respectively.
Among the 46 vibrational modes are 10 Raman-active ones. 
In choosing a basis set for the first-principles calculation, one must make a 
compromise between accuracy and CPU time. 
 Without significant computational cost, one can do B3LYP/STO-3G 
calculation and still obtain results more accurate than any RHF calculations. Going 
beyond STO-3G for B3LYP cases requires a drastic increase in CPU time. Suprisingly, 
going just to 3-21G provides the most accurate results, while for the bigger basis 
set 6-31G, the results are worse and  adding a polarized function to 6-31G only 
slightly improves the results. The 3-21G basis set gives  systematically lower 
frequencies  than the 6-31G basis set, while the frequencies obtained from the 
6-31G* basis set typically lie in between the results of the other two basis sets. 
In contrast, 6-31G* does provide the most accurate bond lengths. This suggests that the 
better accuracy of 3-21G is foirtuitous. Increasing the basis set to 6-31G stiffens the 
bonds, while adding the polarization function compensates by softening the bonds. 
In comparison with the B3LYP results, RHF calculated 
frequencies are too high due to an incorrect description of bond dissociation which 
leads to an increased force constant, while B3LYP's with large basis sets are 
generally in good agreement with the experiments of Wang {\sl et al.}\cite{ir3}. 
In comparison with both effects, we find that the 
correlation effect on the RAFs is stronger than the basis set effect. This 
demonstrates the importance of electron correlation in the accurate description 
of the vibrational frequencies of these molecules. 

\end{multicols}

\noindent
{\bf Table II:}  RHF and B3LYP hybrid DFT calculations
of  Raman scattering activities ( $I_{raman}$, in $10^{-14} {\rm m}^{4}/{\rm kg}$ )
of ${\rm C}_{60}$ with the corresponding vibrational modes and
frequencies $\nu$ (${\rm cm}^{-1}$). ${\rm h}_{\rm g}$ and
${\rm a}_{\rm g}$  modes are unpolarized and polarized, respectively.
Numbers in the parenthesis  are the relative errors between the calculated and 
the  experimental frequencies $\nu^{exp}$ (see Table III) of Wang 
{\sl et al.}\cite{rm1}.  
\begin{center}
\begin{tabular}{cccccccccc}\hline\hline
 & &\multicolumn{2}{c}{\underline{ \ \ \ \ \ \ STO-3G\ \ \ \  }}
&\multicolumn{2}{c}{\underline{\  \ \ \ \ \  \ \ \ 3-21G\ \ \ \ }}
&\multicolumn{2}{c}{\underline{\ \ \  \ \ \  \ \ \  6-31G\ \  \ \ }}
&\multicolumn{2}{c}{\underline{\ \ \ \ \ \ \  \ 6-31G*\ \  \  \ }}\\ 
{\small Method} &{\small Mode}  & ${\small I_{raman}}$ & $\nu$
    & ${\small I_{raman}}$ & $\nu$
    & ${\small I_{raman}}$ & $\nu$
    & ${\small I_{raman}}$ & $\nu$\\ \hline
{\small RHF } & {\small  ${\rm a}_{\rm g}$ } & {\small 3894 } & {\small  1684  (14.6\%) } 
& {\small 3467 } & {\small  1604 (9.2\%) } & {\small 3840 } & {\small   1637 (11.4\%)} 
& {\small  3968 } & {\small  1600 (8.9\%) }  \\
 &  & {\small  441 } & {\small  553 (21.1\%) } & {\small 488 } & {\small  518 (5.0\%) } 
& {\small 558 } & {\small   526 (6.8\%)  } & {\small  574 } & {\small  527  (6.4\%) }  \\
  & {\small  ${\rm h}_{\rm g}$    } & {\small 255 } & {\small 1912 (21.6\%) } & {\small 224 } 
& {\small 1772 (12.7\%) } & {\small 279  } & {\small  1799 (14.4\%) } & {\small  283 } & {\small  1791 (13.9\%) }  \\
  & {\small     } & {\small 34  } & {\small 1658 (16.3\%) } & {\small 35 } & {\small 1546 (8.4\%)  } 
& {\small 38  } & {\small 1585 (11.1\%) } & {\small 22 } & {\small  1562 (9.5\%)  }  \\
  & {\small     } & {\small 188 } & {\small 1482 (18.7\%) } & {\small 215 } & {\small 1326 (6.2\%) } 
&{\small 199 } & {\small  1377 (10.3\%)} & {\small  210 } & {\small  1380 (10.6\%) } \\
  & {\small    } & {\small 87 } & {\small 1290 (17.4\%) } & {\small 76 } & {\small 1184 (7.7\%) } 
& {\small 44 } & {\small  1208 (9.9\%)} & {\small  105 } & {\small  1208 (9.9\%) }  \\
  & {\small     } & {\small 38  } & {\small 886 (14.6\%) } & {\small 38 } & {\small 828 (7.2\%) } 
& {\small 5  } & {\small  843 (9.0\%)} & {\small  46 } & {\small  840 (8.7\%)  }  \\
  & {\small   } & {\small 11  } & {\small 836 (18.1\%) } & {\small 16 } & {\small 761 (7.5\%) } 
& {\small 5  } & {\small  821 (15.9\%)} & {\small  9 } & {\small  794 (12.2\%) }  \\
  & {\small      } & {\small 14  } & {\small  509 (18.0\%)  } & {\small 19 } & {\small 475 (10.2\%) } 
& {\small 16  } & {\small  496 (14.9\%)} & {\small 12 } & {\small  482 (11.9\%)   }  \\
  & {\small     } & {\small  75 } & {\small 302 (11.9\%) } & {\small 111 } & {\small 295 (9.2\%) } 
& {\small 116 } & {\small  296 (9.5\%)} & {\small  115 } & {\small  289 (7.1\%) }  \\
   &  & & &  & & &  & &  \\
{\small B3LYP } & {\small  ${\rm a}_{\rm g}$  } & {\small 3008 } & {\small  1549 (5.4\%) } & {\small 2643   } 
& {\small 1501  (2.1\%) } & {\small 2758 } & {\small  1524 (3.7\%) } & {\small 2760 } & {\small 1504 (2.4\%) }  \\
     & {\small   } & {\small 654 } & {\small  502 (1.9\%)  } & {\small 681   } & {\small   491  (0.4\%)  }
 & {\small 750 } & {\small  496 (0.6\%) } & {\small 724 } & {\small 489 (0.8\%)  }   \\
  & {\small  ${\rm h}_{\rm g}$  } & {\small 320 } & {\small 1677 (6.6\%) } & {\small 297   } 
& {\small   1609 (2.3\%) } & {\small 325  } & {\small  1627 (3.4\%) } & {\small 312 } & {\small 1618 (2.8\%) }   \\
  & {\small   } & {\small 28  } & {\small 1500 (5.2\%) } & {\small 18  } & {\small 1436 (0.7\%) } & {\small 23  } 
& {\small 1466 (2.8\%) } & {\small 19 } & {\small 1455 (2.0\%) } \\
  & {\small   } & {\small 138 } & {\small 1332 (6.7\%) } & {\small 168 } & {\small 1231 (1.4\%) } 
& {\small 151 } & {\small 1274 (2.1\%) } & {\small 170 } & {\small 1275 (2.1\%) }   \\
  & {\small    } & {\small 74 } & {\small 1166 (6.1\%) } & {\small 71   } & {\small   1112 (1.2\%) } 
& {\small 75 } & {\small  1129 (2.8\%) } & {\small 71 } & {\small 1125 (2.4\%) }  \\
  & {\small   } & {\small 47  } & {\small 802 (3.8\%)  } & {\small 48   } & {\small  781 (1.1\%) } 
& {\small 44 } & {\small  788 (1.9\%) } & {\small 48 } & {\small 766 (0.9\%) }  \\
  & {\small   } & {\small 4  } & {\small 734 (3.7\%)  } & {\small 5   } & {\small  678 (4.3\%)  } 
& {\small 4 } & {\small  738 (4.3\%) } & {\small 5 } & {\small 718 (1.4\%) }  \\
  & {\small   } & {\small 4  } & {\small 449 (4.2\%)  } & {\small 5   } & {\small 429 (0.5\%) } 
& {\small 4 } & {\small  448 (3.9\%) } & {\small 5 } & {\small 436 (1.2\%) }  \\
{\small } & {\small   } & {\small 95 } & {\small 271 (0.5\%) } & {\small 128 } & {\small  271 (0.3\%) } 
& {\small 135 } & {\small   272 (0.2\%) } & {\small 130 } & {\small 266 (1.7\%) }  \\ \hline
\end{tabular}
\end{center}

\begin{multicols}{2}

For comparison, Table III lists the calculated vibrational frequencies
of $C_{60}$ obtained by using various theories, for example, the semi-empirical
MNDO\cite{mndo88} and  QCFF/PI\cite{negri88} methods.  Of
these,  the QCFF/PI method, which has been
 parameterized mainly with respect to vibrational frequencies of conjugated and aromatic
hydrocarbons\cite{warshel72},
results in the best results although it gives less satisfactory
geometry. Such accurate
prediction implies that the electronic structures of $C_{60}$ is
not much different from other
aromatic hydrocarbons\cite{negri88}.
H\"{a}ser {\sl et al.} \cite{haser91} showed that
the approximate harmonic frequencies for the two $a_{g}$
vibrational modes of ${\rm C}_{60}$  are 1615 ${\rm cm}^{-1}$ (9.9\%) and
487 ${\rm cm}^{-1}$ (1.2\%) at HF/DZP,  1614 ${\rm cm}^{-1}$ (9.9\%) and
483  ${\rm cm}^{-1}$ (2.0\%) at HF/TZP,  1614 (9.8\%) ${\rm cm}^{-1}$ and
437 (12.2\%) ${\rm cm}^{-1}$ at MP2/DZP, and  1586 (7.9\%) ${\rm cm}^{-1}$ and
437 (12.2\%)  ${\rm cm}^{-1}$ at MP2/TZP, where the percentages  in the parenthesis are the relative errors
of the calculated results to the experimental frequencies obtained by Wang {\sl et al.}\cite{rm1}. Their HF calculations are in
agreement with our RHF/3-21G results. Their MP2 results are more accurate when
obtained with large basis sets, which also demonstrates the importance
of electron correlation in predicting accurately the vibrational frequencies
of a molecule.

\end{multicols}

\noindent
{\bf Table III:} The vibrational frequencies obtained by other theoretical
calculations performed by Choi {\sl et al.}\cite{choi00a}, Bohnen {\sl et al.}\cite{bohnen95},
 Stanton {\sl et al.}\cite{mndo88}, Negri {\sl et al.}\cite{negri88},  Jishi {\sl et al.}\cite{fcm92},
  Dixon {\sl et al.} \cite{dad95}, Hara {\sl et al.}\cite{hara01} and  Onida {\sl et al.}\cite{onida94}.
Numbers in the parenthesis are the relative errors to the
experimental frequencies, $\nu^{exp}$, of Wang {\sl et al.}\cite{rm1}.
\begin{center}
\begin{tabular}{llllllllll}\hline\hline
{\small SIFC}  & {\small LDA} & {\small LDA} & {\small LDA} & {\small CPMD} &{\small QCFF/PI} & {\small MFCM}  &{\small MNDO}   & {\small Exp.} \\
{\small \cite{choi00a}}  &{\small \cite{hara01}}  & {\small \cite{dad95}} &{\small\cite{bohnen95}} &{\small  \cite{onida94}}
      &{\small Ref.\cite{negri88}} &{\small \cite{fcm92}} & {\small \cite{mndo88}}  & {\small \cite{rm1}}\\ \hline
$a_{g}$ & & & & & & & & & \\
{\small 1474 (0.3\%)}  &{\small  1531 (4.2\%)} & {\small 1525 (3.8\%)} &{\small  1475 (0.4\%)} &{\small  1447 (1.5\%)}   
&{\small  1442 (1.8\%)} & {\small  1468 (0.1\%) }& {\small 1667 (13.5\%)} & 1469 \\
{\small 484 (1.8\%)} &{\small  502 (1.8\%) } &{\small  499 (1.2\%) } &{\small  481 (2.4\%) } &{\small  482 (2.2\%) }  
 &{\small  513 (4.1\%)}  &{\small  492 (0.2\%) }&{\small  610  (23.7\%)} & 493  \\
  & & & & & & & & & \\ 
$h_{g}$ & & & & & & & & & \\ 
 {\small 1582 (0.6\%)}  &{\small  1609 (2.3\%)} &{\small  1618 (2.8\%)} &{\small  1580 (0.4\%)} &{\small  1573 (0.0\%)}  
 &{\small  1644 (4.5\%)} &{\small  1575 (0.1\%)} & {\small  1722  (9.5\%)} & 1573 \\
 {\small  1419 (0.5\%)}         &{\small  1475 (3.4\%)} &{\small  1475 (3.4\%)} &{\small  1422 (0.3\%)} &{\small  1394 (2.2\%)}   
& {\small 1465 (2.7\%)} & {\small  1401 (1.8\%)} &{\small  1596  (11.9\%)} & 1426  \\
 {\small 1250 (0.2\%)}   &{\small  1288 (3.2\%)} &{\small  1297 (3.9\%)} &{\small  1198 (4.0\%)} &{\small  1208 (3.2\%)}   
&{\small  1265 (1.4\%)} &{\small  1217 (2.5\%)}& {\small 1407  (12.7\%)}   & 1248    \\
 {\small 1117 (1.6\%)}          &{\small  1129 (2.7\%)} &{\small  1128 (2.6\%)} &{\small  1079 (1.8\%)} &{\small  1098 (0.1\%)}  
 &{\small  1154 (5.0\%)} & {\small  1102 (0.3\%)} &{\small  1261  (14.7\%)}  & 1099   \\
  {\small  782 (1.2\%)}         &{\small  794 (2.7\%) } &{\small  788 (1.9\%)}  &{\small  763 (1.3\%)}  &{\small   775 (0.3\%)}   
&{\small 801 (3.6\%)} &{\small  788 (1.9\%)} &{\small  924  (19.5\%)} & 773   \\
  {\small  704 (0.6\%)}         &{\small  711 (0.4\%) } &{\small  727 (2.7\%)}  &{\small  716 (1.1\%)}  &{\small  730 (3.1\%)}   
 &{\small 691 (2.4\%)} &{\small  708 (0.0\%)} & {\small 771  (8.2\%)}  &  708     \\
  {\small 436 (1.2\%)}         &{\small  430 (0.2\%) } &{\small  431 (0.0\%)}  &{\small  422 (2.1\%)}  &{\small  435 (0.9\%)}   
&{\small 440 (2.1\%)} &{\small  439 (1.9\%)} & {\small  447  (3.7\%)} & 431      \\
  {\small 272 (0.7\%)}  &{\small  269 (0.4\%) } &{\small  261 (3.3\%)}  &{\small  263 (2.6\%)}  &{\small   261 (3.3\%)}  
&{\small  258 (4.4\%)} &{\small  269 (0.4\%)} &{\small  263  (2.7\%)} & 270 \\ \hline
\end{tabular}
\end{center}

\begin{multicols}{2} 
 
In addition,  a number of force-constant models (FCMs) that 
include  the interactions up to the second-nearest neighbors\cite{wu87,cyvin88,weeks89} are 
used to calculate the vibrational frequencies of $C_{60}$. None of them yield good agreement with
the experiment data.  For example, an empirical force field, which
has been parameterized with respect to polycyclic aromatic hydrocarbons,
is used with the H\"{u}ckel theory and predicts that the vibrational frequencies
of the two $a_{g}$ modes are 1409 ${\rm cm}^{-1}$ and  388
${\rm cm}^{-1}$\cite{cyvin88}, which are too low. However, the
modified FCM (MFCM) by Jishi {\sl et al.}\cite{fcm92}  considered interactions
up to the third-nearest neighbors. The results obtained by using MFCM 
\cite{fcm92}, as shown in Table III, are in excellent agreement with the experiments 
of  Wang {\sl et al.}\cite{ir3}.
 
Table III also list the calculated vibrational frequencies of $C_{60}$ 
obtained by other DFT methods, for example, local density approximation (LDA)
\cite{bohnen95,dad95,hara01} and DFT-LDA-based Car-Parrinello 
molecular dynamics (CPMD) simulation\cite{onida94}. In general, those calculated results
are in good agreement with experiment.

Recently,  Choi {\sl et al.}\cite{choi00a} have performed B3LYP vibrational calculations 
of ${\rm C}_{60}$ with a 3-21G basis set but involving scaling of the internal 
force constants (SIFC)  $\tilde{K}_{ij}^{int}$ by using Pulay's method\cite{pulay}, i.e.,
\begin{equation}
\tilde{K}_{ij}^{scaled} = (s_{i}s_{j})^{1/2}\tilde{K}_{ij}^{int}, 
\end{equation}
where $\tilde{K}_{ij}^{int}$ is the force constant in internal coordinates ( the Gaussian 98 program 
\cite{gaussian} uses this form),  and $s_{i}$ and $s_{j}$ are scaling factors for the $i$th and $j$th 
redundant internal coordinates, respectively. They optimized the scaling factors by minimizing 
the root-mean-square deviations between the experimental and calculated scaled frequencies. Their 
results are also listed in Table III. Their scaling
procedure does improve the accuracy for the 10 Raman-active vibrational
frequencies of ${\rm C}_{60}$.

It should be mentioned that although vibrational calculations are all done for 
an isolated ${\rm C}_{60}$ molecule, the calculated results are usually compared to experimental data on 
${\rm C}_{60}$ molecular vibrations in the solid state, in the gas-phase or in  solution. The theoretical justification for this 
lies in the weak Van der Waals bonds between ${\rm C}_{60}$ molecules\cite{book1}. The comparison between 
the observed vibrational spectra of ${\rm C}_{60}$\cite{book1} demonstrated that the measured 
mode frequencies are slightly different from one phase to another, showing that the interaction 
between ${\rm C}_{60}$ molecules is weak. 
 
In Table IV and V, we list the RAFs for
${\rm C}_{48}{\rm N}_{12}$ calculated with RHF and B3LYP hybrid DFT methods 
, respectively, and different basis sets.  In contrast with $C_{60}$, we find that
there are in total 116 different vibrational modes\cite{xiecpl02} for 
${\rm C}_{48}{\rm N}_{12}$ because of its lower symmetry, $S_{6}$\cite{xiecpl02}. These vibrational 
modes are classified in 58
doubly-degenerate and 58 nondegenerate modes. Among those vibrational
modes, there are 58 IR-active vibrational modes\cite{xiecpl02}  and 
58 Raman-active modes including 29 doubly-degenerate and 29 non-degenerate ones as 
listed in Table IV and V.  Similar to ${\rm C}_{60}$,  including the electron correlation 
and  increasing the basis size leads to a redshift of the RAFs of  ${\rm C}_{48}{\rm N}_{12}$. 
From our calculated results,  we find that the nitrogen-substitutional doping results in a symmetry  
lowering and an increase in the reduced mass. The symmetry lowering splitts some of 
the degenerate vibrational modes observed in $C_{60}$ and  makes many more modes  Raman-active for
 ${\rm C}_{48}{\rm N}_{12}$. Overall,  the increase of the reduced mass  red-shifts  the vibrational 
frequencies of  $C_{60}$.

\subsection{Raman Scattering Activities}
 
 In this paper, we only calculate non-resonant
Raman intensities. We performed  {\sl ab initio} calculations of Raman scattering activities
$I_{raman}$ \cite{book8a,book8b}  for the optimized geometries of ${\rm C}_{48}{\rm N}_{12}$
and ${\rm C}_{60}$ by using the Gaussian 98 program
\cite{gaussian,nist} with RHF and B3LYP hybrid DFT methods.

\vspace{3cm}

\noindent
{\bf Table IV}:  Fifty eight Raman-active frequencies 
$\nu$ (${\rm cm}^{-1}$)for ${\rm C}_{48}{\rm N}_{12}$ calculated 
by using RHF method with STO-3G, 3-21G, 6-31G and  6-31G*  basis sets.  
\begin{center}
\begin{tabular}{cccc|cccc}\hline\hline
\multicolumn{4}{c|}{\small Doubly-degenerate Modes} &
\multicolumn{4}{|c}{\small Non-degenerate Modes} \\ \hline
{\small STO-3G} & {\small 3-21G} & {\small 6-31G} & {\small 6-31G*}  & 
{\small STO-3G} & {\small 3-21G} & {\small 6-31G} & {\small 6-31G*} \\ \hline
{\small 262.9} & {\small  262.1 } & {\small  268.3} & {\small  260.7 } 
 &{\small 293.9} & {\small  287.6} & {\small  291.2} & {\small 287.9 }\\ 
 {\small   285.8} & {\small 280.9} & {\small  285.5} & {\small 280.8} 
 & {\small 444.6} & {\small 406.1} & {\small  428.9} & {\small  387.5}\\ 
 {\small   429.1} & {\small 408.9} & {\small  427.5} & {\small  412.9} 
& {\small  462.5} & {\small 415.1} & {\small  447.6} & {\small  424.5}\\ 
{\small   470.4} & {\small 433.3} & {\small  454.6} & {\small  440.1} 
& {\small  510.2} & {\small 455.2} & {\small  483.2} & {\small 454.0}\\ 
 {\small  513.6} & {\small 478.7} & {\small  496.7} & {\small  488.3} 
 & {\small  522.6} & {\small 490.8} & {\small  511.4} & {\small  488.6}\\
 {\small  583.2} & {\small 516.9} & {\small  548.6} & {\small  539.9} 
 & {\small  575.6} & {\small 520.8} & {\small  542.8} & {\small  523.7}\\  
 {\small   603.5} & {\small 609.5} & {\small  611.3} & {\small  592.2} 
 & {\small  608.3} & {\small 566.1} & {\small  586.8} & {\small  575.2}\\ 
 {\small  651.8} & {\small 647.3} & {\small  649.5} & {\small  628.6} 
 & {\small  645.7} & {\small 637.1} & {\small  639.7} & {\small 614.7} \\
 {\small  725.4} & {\small 698.9} & {\small  736.4} & {\small  706.2} 
& {\small  650.4} & {\small 650.6} & {\small  652.2} & {\small 636.8 }\\ 
 {\small   773.4} & {\small 726.4} & {\small  759.2} & {\small  740.9} 
 & {\small  723.1} & {\small 654.8} & {\small   670.3} & {\small  651.8}\\ 
 {\small  780.2} & {\small 761.6} & {\small   785.1} & {\small  757.4} 
 & {\small  744.2} & {\small 705.2} & {\small   748.8} & {\small  716.4}\\ 
 {\small   831.2} & {\small 775.8} & {\small  806.9} & {\small  788.0} 
 & {\small  777.3} & {\small  759.7} & {\small  783.6} & {\small  738.3}\\ 
 {\small   848.0} & {\small 820.3} & {\small  833.4} & {\small  820.8} 
 & {\small  831.6} & {\small 778.6} & {\small  802.4} & {\small 772.8}\\ 
 {\small   870.7} & {\small 843.6} & {\small  852.5} & {\small  840.1} 
 & {\small  866.8} & {\small  824.8} & {\small  836.1} & {\small  819.0} \\ 
 {\small  934.6} & {\small  942.6} & {\small  943.5} & {\small  917.9} 
 & {\small  923.6} & {\small  918.4} & {\small  919.0} & {\small  897.0}\\ 
 {\small  961.3} & {\small  974.4} & {\small  965.8} & {\small  932.8} 
 & {\small 952.9} & {\small  964.6} & {\small  955.7} & {\small  916.2}\\ 
 {\small  1238.4} & {\small 1159.9} & {\small 1191.9} & {\small  1164.4} 
 & {\small  1216.2} & {\small 1164.9} & {\small 1192.3} & {\small  1174.7}\\ 
 {\small  1280.9} & {\small 1196.7} & {\small 1225.0} & {\small  1214.5} 
 & {\small  1327.5} & {\small 1243.7} & {\small 1264.1} & {\small 1251.1} \\ 
 {\small  1313.3} & {\small 1238.4} & {\small 1261.7} & {\small  1249.9} 
& {\small  1344.1} & {\small 1259.7} & {\small 1281.3} & {\small 1272.8} \\ 
 {\small  1422.9} & {\small 1276.3} & {\small 1324.6} & {\small  1317.6} 
 & {\small  1400.2} & {\small 1335.0} & {\small 1374.0} & {\small 1347.6} \\    
 {\small  1476.0} & {\small 1340.6} & {\small 1382.7} & {\small  1377.4} 
 & {\small  1468.7} & {\small 1347.1} & {\small 1389.0} & {\small  1376.6}\\   
 {\small  1546.2} & {\small 1420.7} & {\small 1456.9} & {\small  1448.6} 
 & {\small  1498.2} & {\small 1374.6} & {\small 1424.4} & {\small 1380.0} \\    
 {\small  1608.0} & {\small 1468.3} & {\small 1517.1} & {\small  1503.4} 
 & {\small  1533.9} & {\small 1414.9} & {\small 1468.7} & {\small  1429.1} \\
 {\small  1647.0} & {\small 1490.0} & {\small 1537.2} & {\small  1540.1} 
 & {\small  1583.4} & {\small 1436.5} & {\small 1485.1} & {\small  1475.9}\\     
 {\small  1716.5} & {\small 1541.4} & {\small 1591.4} & {\small  1586.5} 
 & {\small  1655.1} & {\small 1527.5} & {\small 1578.0} & {\small  1551.9}\\   
 {\small  1780.9} & {\small 1631.3} & {\small 1670.8} & {\small  1663.5} 
 & {\small  1721.2} & {\small 1559.5} & {\small 1612.3} & {\small  1602.1}\\ 
 {\small  1831.0} & {\small 1669.5} & {\small 1709.9} & {\small  1719.9} 
& {\small  1766.4} & {\small 1612.4} & {\small 1648.6} & {\small 1625.7} \\   
 {\small  1882.8} & {\small 1707.7} & {\small 1749.3} & {\small  1758.7} 
 & {\small  1817.4} & {\small 1643.5} & {\small 1686.6} & {\small  1680.1}\\   
 {\small  1915.5} & {\small 1735.1} & {\small 1777.7} & {\small  1788.1} 
& {\small  1946.3} & {\small 1766.7} & {\small 1803.8} & {\small 1808.7} \\ \hline
\end{tabular}
\end{center}

\noindent
{\bf Table V}: {\small Fifty eight Raman-active frequencies 
$\nu$ (${\rm cm}^{-1}$)for ${\rm C}_{48}{\rm N}_{12}$ calculated 
by using B3LYP hybrid DFT methods with STO-3G, 3-21G, 
6-31G and  6-31G*  basis sets. } 
\begin{center}
\begin{tabular}{cccc|cccc}\hline\hline
\multicolumn{4}{c|}{\small Doubly-degenerate Modes} &
\multicolumn{4}{|c}{\small Non-degenerate Modes} \\ \hline
{\small STO-3G} & {\small 3-21G} & {\small 6-31G} & {\small 6-31G*}  & 
{\small STO-3G} & {\small 3-21G} & {\small 6-31G} & {\small 6-31G*} \\ \hline
{\small 245.1 } & {\small 247.7} & {\small   252.1} &{\small 245.1} 
& {\small   263.8 } & {\small 264.3} & {\small 267.5} & {\small 263.6} \\ 
{\small 258.9 } & {\small  261.1} & {\small  264.5} &{\small 259.5 }  
& {\small 381.9 } & {\small  368.2} & {\small  389.4} &{\small 376.2 } \\ 
{\small 376.1 } & {\small  367.7} & {\small  386.7} &{\small 371.2 }  
& {\small  405.7 } & {\small 397.7} & {\small 415.0} &{\small 398.3 } \\ 
{\small 410.2 } & {\small  388.2} & {\small  408.9} &{\small 396.0}  
& {\small 444.4 } & {\small  415.4} & {\small 441.7} &{\small 423.8 } \\ 
{\small  446.7 } & {\small   428.7} & {\small 446.2} &{\small 436.5  }  
& {\small 471.2 } & {\small 458.4} & {\small  477.3}  &{\small 466.6  } \\
{\small 492.6 } & {\small  449.5} & {\small   482.0} &{\small 471.6 }  
& {\small 498.3 } & {\small  490.5} & {\small 500.2} &{\small 494.9 } \\  
{\small 548.8 } & {\small  567.5} & {\small  566.0} &{\small 551.1 }  
& {\small 544.1 } & {\small  504.5} & {\small 544.2} &{\small 510.0} \\ 
{\small 580.3 } & {\small   596.3} & {\small  596.8} &{\small 580.6 }  
& {\small 577.5 } & {\small 588.2} & {\small  592.1} &{\small 575.7}  \\
{\small 617.3 } & {\small   612.8} & {\small  652.2} &{\small 626.6 }  
& {\small 585.2 } & {\small 597.8} & {\small  601.5} &{\small 588.1 } \\ 
{\small 649.4} & {\small   629.1} & {\small  662.4} &{\small 641.2 }  
& {\small 621.2} & {\small 603.0} & {\small  616.3} &{\small 596.9} \\ 
{\small 665.2} & {\small   670.6} & {\small  698.1} &{\small 671.1 } 
& {\small 624.9} & {\small  614.1} & {\small 652.1} &{\small 627.0} \\ 
{\small 715.9} & {\small   695.0} & {\small  713.9} &{\small 698.6 }  
& {\small 657.8} & {\small  657.1} & {\small 684.2} &{\small 656.2} \\ 
{\small 766.6} & {\small   760.4} & {\small  768.0} &{\small 765.8 }  
& {\small  729.5} & {\small  690.1} & {\small 709.4} &{\small 688.5} \\ 
{\small 781.7} & {\small   776.0} & {\small  782.4} &{\small 779.7 }  
& {\small 768.2} & {\small 764.8} & {\small  772.5} &{\small 765.5}  \\ 
{\small 840.8} & {\small   865.3} & {\small  860.2} &{\small 842.6}  
& {\small 835.5} & {\small 850.5} & {\small  844.4} &{\small 829.7} \\ 
{\small 858.1} & {\small   887.4} & {\small  876.6} &{\small 853.9}  
& {\small  855.9} & {\small  881.0} & {\small 868.6} &{\small 843.7} \\ 
{\small 1123.7} & {\small   1084.4} & {\small 1104.9} &{\small 1092.7}  
& {\small 1114.7} & {\small 1079.8} & {\small 1099.8} &{\small 1092.1} \\ 
{\small 1162.2} & {\small   1122.8} & {\small 1140.6} &{\small 1138.0 } 
& {\small 1193.6} & {\small 1160.0} & {\small 1172.5} &{\small 1161.6}  \\ 
{\small 1181.5} & {\small   1146.8} & {\small 1162.9} &{\small 1155.8 }  
& {\small 1220.7} & {\small 1181.4} & {\small 1195.8} &{\small 1189.0}  \\ 
{\small 1274.1} & {\small   1186.0} & {\small 1220.0} &{\small 1222.6 } 
& {\small 1315.9} & {\small 1240.9} & {\small 1267.3} &{\small 1265.1}  \\    
{\small 1318.1} & {\small   1243.4} & {\small 1273.6} &{\small 1271.7 }  
& {\small 1337.1} & {\small 1253.2} & {\small 1280.2} &{\small 1279.0} \\   
{\small 1386.2} & {\small   1310.8} & {\small 1336.6} &{\small 1336.3 }  
& {\small 1366.7} & {\small 1292.5} & {\small 1322.1} &{\small 1319.8} \\    
{\small 1451.9} & {\small   1369.3} & {\small 1406.8} &{\small 1404.1 }  
& {\small 1416.4} & {\small 1346.6} & {\small 1382.6} &{\small 1378.9}  \\
{\small 1466.1} & {\small   1387.9} & {\small 1420.4} &{\small 1419.7 }  
& {\small 1438.7} & {\small 1372.8} & {\small 1414.5} &{\small 1397.4} \\     
{\small 1503.9} & {\small   1415.3} & {\small 1451.6} &{\small 1451.0 }  
& {\small 1481.0} & {\small 1418.8} & {\small 1453.2} &{\small 1441.1} \\   
{\small 1570.8} & {\small   1488.6} & {\small 1516.8} &{\small 1515.4 }  
& {\small 1530.7} & {\small 1440.4} & {\small 1477.2} &{\small 1476.6} \\ 
{\small 1595.2} & {\small   1516.0} & {\small 1550.6} &{\small 1551.0 }  
& {\small 1558.4} & {\small 1487.1} & {\small 1511.4}  &{\small 1505.3} \\   
{\small 1633.1} & {\small   1544.0} & {\small 1577.6} &{\small 1577.8 } 
& {\small 1580.5} & {\small 1508.5} & {\small 1545.1} &{\small 1530.4} \\   
{\small 1661.7} & {\small   1577.6} & {\small 1609.8} &{\small 1609.8}  
& {\small 1682.7} & {\small 1592.0} & {\small 1623.7} &{\small 1622.9} \\ \hline
\end{tabular}
\end{center}

To test the basis set dependence of the theoretical RSAs, calculations for
basis sets of different sizes  have been performed for ${\rm C}_{60}$.
In Table II are listed the  RSAs for ${\rm C}_{60}$ calculated with STO-3G, 3-21G, 6-31G and 6-31G*.
For each basis set, the strongest Raman-active lines of  ${\rm C}_{60}$ are the two ${\rm a}_{\rm g}$
modes, which is consistent with  Raman experiments
\cite{ir2,ir3,ir4,rm1,rm2}. The
${\rm a}_{\rm g}$ modes are identified  by their
polarized character, which strongly suggests that both ${\rm a}_{\rm g}$
modes are totally symmetric\cite{dress92}. The remaining 8 Raman-active
modes are unpolarized,  consistent with the fivefold-degenerate
${\rm h}_{\rm g}$ symmetry\cite{dress92}.  Also, we find that
electron correlation and basis set  change significantly
the RSA of a given Raman-active mode. However,  the correlation effect is
stronger than the basis set effect. In general,  electron correlation
 reduces the RSA of the strongest Raman-active $a_{g}$ mode of
${\rm C}_{60}$ by about 25\%. For other specific Raman-active modes, the electron correlation
predicts at least a 20\% decrease (or increase) of the RSA  obtained by the RHF method.
Increasing the size of the basis set from STO-3G to 6-31G results in about a 10\%
decrease (or increase) of RSA, while adding polarization functions leads to
about a 3\% decrease (or increase) of RSA.  For a specific mode (for example, the $q$th one),
we find that the $k$th atomic displacement  $\xi_{kq}$ of this mode changes little
as we choose different basis sets or/and consider the electron correlation.
Thus, according to Eq.(5), the RSA mainly depends on the derivative of the polarizability
$\alpha$ with respect to atomic coordinates $R$, which are related to the choice of
basis sets and the inclusion of electron correlations.

\begin{center}
\epsfig{file=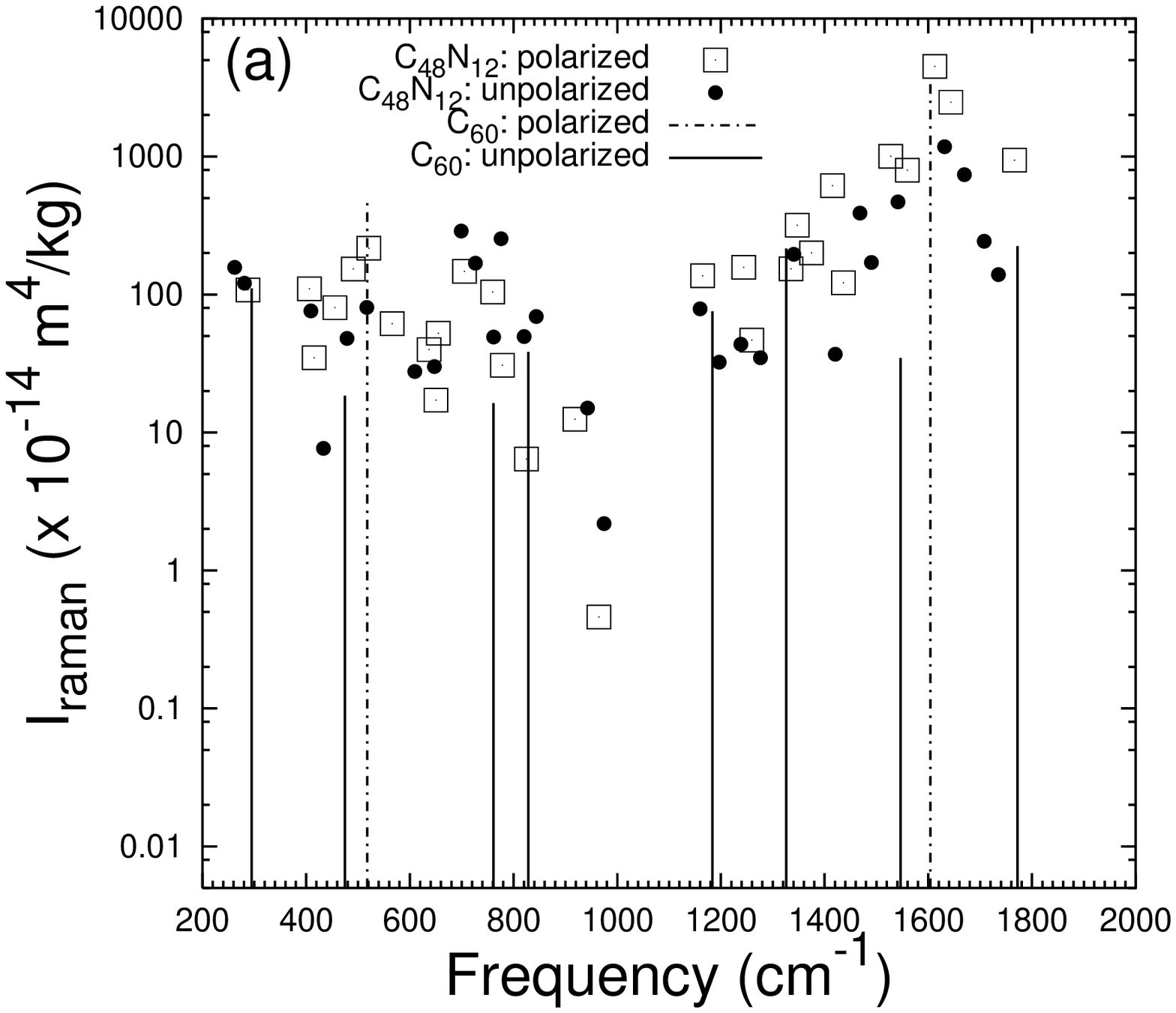,width=7cm,height=6cm}
 
\vspace{0.5cm}
 
\epsfig{file=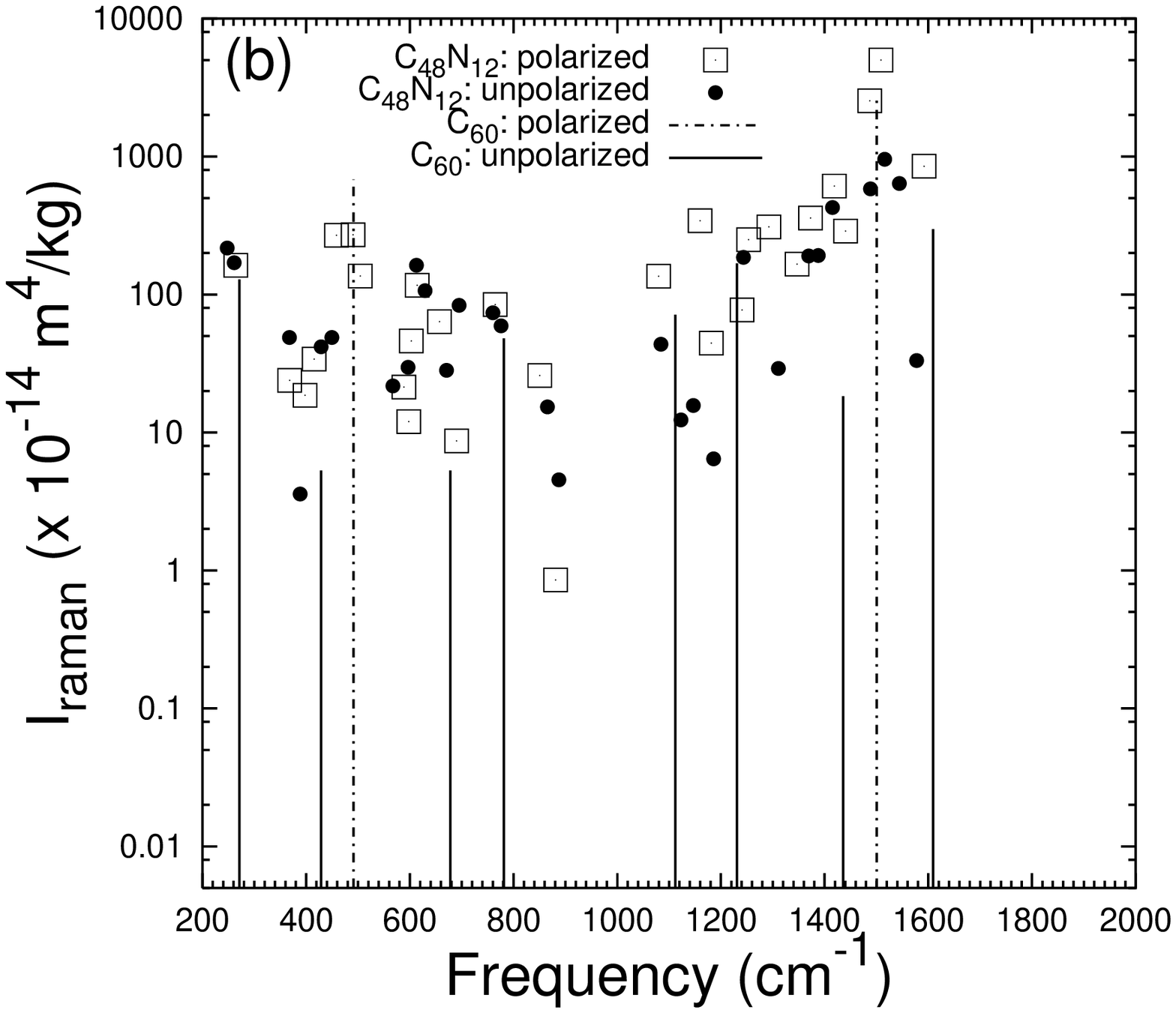,width=7cm,height=6cm}
 
\end{center}

\begin{quote} 
{\bf FIG.2}:  {\small {\sl Ab initio} calculations of Raman
scattering activities ($I_{raman}$, in $10^{-14} {\rm m}^{4}/{\rm kg}$) in
${\rm C}_{48}{\rm N}_{12}$ with (a) RHF/3-21G and (b) B3LYP/3-21G. Open squares and  filled circles are non-degenerate polarized and
doubly-degenerate unpolarized Raman-active modes, respectively for
${\rm C}_{48}{\rm N}_{12}$. The solid and dot-dashed lines are the
calculated unpolarized and polarized Raman spectral lines of
${\rm C}_{60}$, respectively.}
\end{quote}

\vspace{2cm}
 
\begin{center}
\epsfig{file=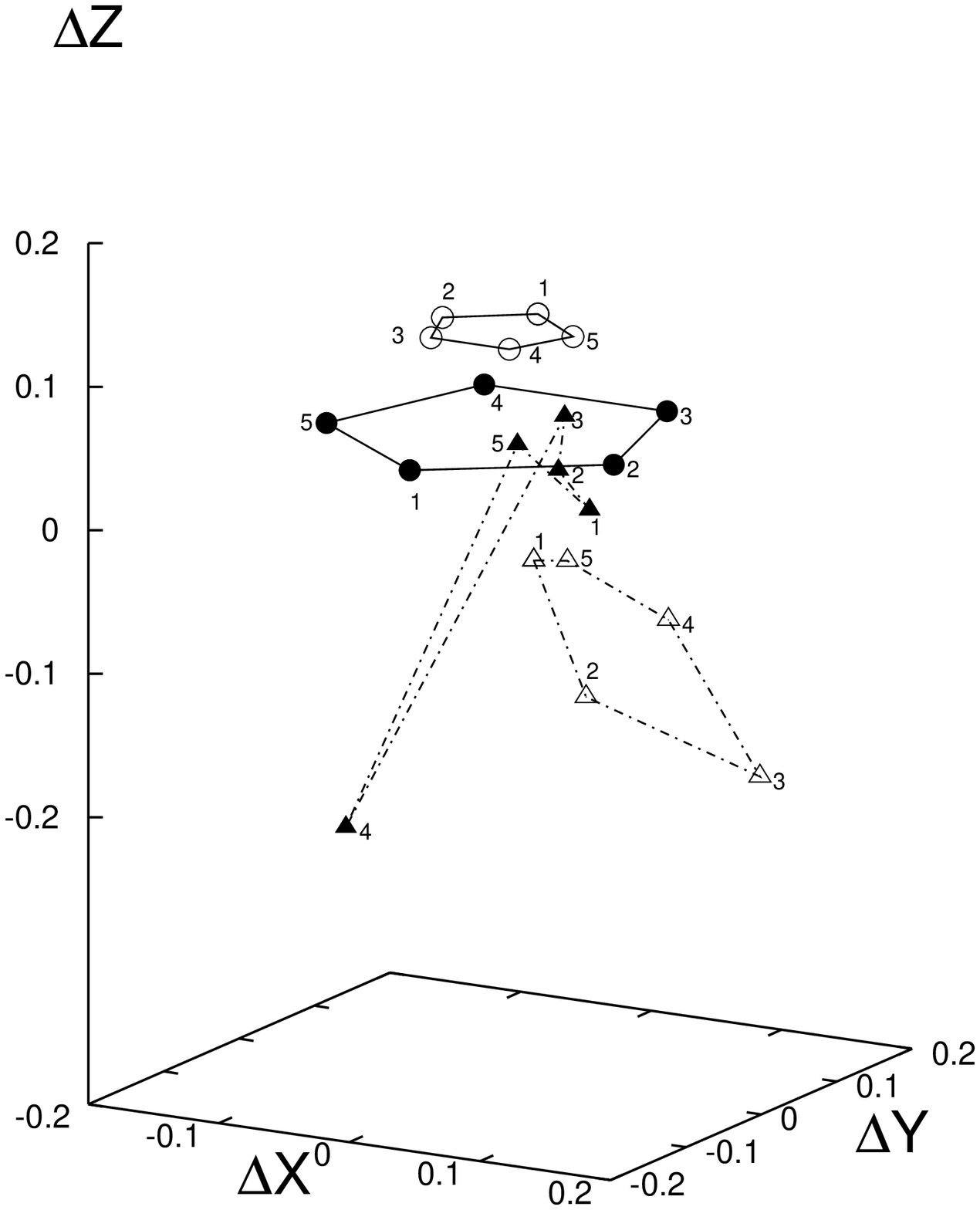,width=7cm,height=6cm}
\end{center}
\begin{quote}
 {\bf FIG.3}: The vibrational displacements of sites 1 to 5 for the strongest Raman spectral lines in
both low- and high-frequency regions for the B3LYP/3-21G case. The open and filled circles are for the
low- and high-frequency cases of ${\rm C}_{60}$, respectively. The open and filled triangles are for the
low- and high-frequency cases of ${\rm C}_{48}{\rm N}_{12}$, respectively.
\end{quote}

In Fig.2(a)(b), we show the calculated RSAs $I_{raman}$ at  the corresponding
RAFs $\nu$ for ${\rm C}_{48}{\rm N}_{12}$ by using RHF/3-21G and B3LYP/3-21G, 
respectively.  We find that the  Raman spectrum of ${\rm C}_{48}{\rm N}_{12}$
separates into  two regions, i.e., high-frequency  (1100 ${\rm cm}^{-1}$ to
1700 ${\rm cm}^{-1}$) and low-frequency
(200 ${\rm cm}^{-1}$ to 1000 ${\rm cm}^{-1}$) regions, which are similar
to those of ${\rm C}_{60}$. In detail,  this aza-fullerene, unlike
${\rm C}_{60}$,  has an equal number of polarized and unpolarized
Raman-active modes in each region and, in particular, has 6 more
Raman-active modes in the low-frequency region than in the high-frequency one.
The strongest Raman spectral lines in both low- and high-frequency regions 
are the non-degenerate polarized ones located at 1509 ${\rm cm}^{-1}$ and  491 ${\rm cm}^{-1}$, 
respectively, which are almost the same as those of the strongest ${\rm a}_{\rm g}$ modes of 
${\rm C}_{60}$. Taking B3LYP/3-21G calculations as an example, we show in Fig.3 the 
vibrational displacements of sites 1 to 5 for  the strongest Raman 
modes in both ${\rm C}_{48}{\rm N}_{12}$ and ${\rm C}_{60}$. For ${\rm C}_{60}$, the pentagon structure 
for the low-frequency case contracts largely in the x-y plane and shows strong collective z-axis vibrations, while 
 the pentagon structure for the high-frequency case expands largely in the x-y plane and shows small collective z-axis
vibrations. In contrast, as a result of N doping, the pentagon structure for ${\rm C}_{48}{\rm N}_{12}$ 
for the low-frequency case expands slightly and shows collective vibration along the 
z-x direction, while the pentagon structure for  
the high-frequecny case  contracts, accompanying a large stretching of  site 4.  
As shown in Fig.2, the lowest and highest RAFs of ${\rm C}_{48}{\rm N}_{12}$  are almost the same 
as those of ${\rm C}_{60}$. Comparing Fig.2(a) with Fig.2(b) shows that the RAFs  of ${\rm C}_{48}{\rm N}_{12}$ predicted 
by RHF method,
as discussed above, are redshifted due to electron correlations.
The effects of basis set and  electron correlations on RSAs of ${\rm C}_{48}{\rm N}_{12}$ 
 are similar to those obtained for ${\rm C}_{60}$. Detailed analysis of Raman-active 
vibrational modes of ${\rm C}_{48}{\rm N}_{12}$  shows that (i) the Raman spectra in the high-frequency region 
($>1400 {\rm cm}^{-1}$) comes mainly from carbon-carbon vibrations; 
(ii) the strong nitrogen-carbon vibrations occur only in the low-frequency region; 
(iii) the  strongest and weakest Raman signals come from the contributions of 
carbon-carbon vibrations.

\section{summary}
 
In summary, we have performed large scale {\sl ab initio} calculations of
RSAs and RAFs  in ${\rm C}_{48}{\rm N}_{12}$ as well as ${\rm C}_{60}$ using B3LYP hybrid DFT  and
RHF methods. We predict that ${\rm C}_{48}{\rm N}_{12}$ has 29
non-degenerate polarized  and 29 doubly-degenerate unpolarized 
RAFs, and the RAF of the strongest spectral lines in the low- and high-
frequency regions and the lowest and highest RAFs are almost the 
same as those of ${\rm C}_{60}$. The 10 RAFs of ${\rm C}_{60}$ 
calculated with the B3LYP hybrid DFT method and large basis set 
are in excellent agreement with experiment. Our study of 
${\rm C}_{60}$ reveals the importance of
electron correlations and basis sets in the {\sl ab initio} calculations,
and our calculated Raman results of ${\rm C}_{60}$ demonstrate the desirable
efficiency and accuracy of the B3LYP hybrid DFT method for obtaining
important quantitative information on the vibrational properties of  
these fullerenes. Finally, we hope that these calculations will provide an 
incentive for Raman measurements on ${\rm C}_{48}{\rm N}_{12}$.

\section*{Acknowledgement}
 
We  thank Dr. Denis A. Lehane  and Dr. Hartmut Schmider for their
technical help. One of us (R.H.X) thanks the High Performance 
Computing Virtual Laboratory  (HPCVL) at Queen's University
for the use of its parallel supercomputing facilities.

\end{multicols}


\begin{references}
\bibitem{nist} Use of this software does not constitute an endorsement
or certification by NIST.
\end{references}
\end{document}